\documentclass[%
 reprint,
showpacs,
 amsmath,amssymb,
 aps,
]{revtex4-1}

\usepackage{fancyhdr}
\usepackage{graphicx}
\usepackage{dcolumn}
\usepackage{bm}

\begin{document}

\title{Insurgent Metamaterials:  light transmission  by forbidden transitions}%
\author{Evgenii E. Narimanov}
\affiliation{School of Electrical and Computer Engineering and Birck Nanotechnology Center, Purdue University, West Lafayette, Indiana 47907, USA}
\date{\today}

\begin{abstract}
We  introduce the concept of the ``insurgent metamaterial'' -- which is a hyperbolic medium that contains quantum emitters with a forbidden optical transition. We show that the resulting electromagnetic response of the composite is dramatically different from that expected from the ``conventional'' hyperbolic medium, and discuss the experimental manifestations of the predicted effect.
\end{abstract}

\maketitle

\section{Introduction: the concept of the Insurgent Metamaterial}

In the picture formed by the conventional wisdom, \cite{BornWolf} the ``forbidden'' transitions  \cite{Boyd}   -- for which the dipole matrix element is exactly zero due to the symmetry of the corresponding wavefunctions \cite{LandauQM} -- usually don't contribute.\cite{Ziman} While there exists (limited) literature on the contributions of such forbidden transition in extreme regimes,\cite{Kaminer} these are usually only observed under strong fields and only lead to small  corrections.

This ``prohibition" that until now has been the "law of the land" in classical optics, \cite{BornWolf}  generally 
 relies on the dipole approximation \cite{Boyd} -- which only applies to the electromagnetic fields that slowly varying in space. However, as the whole reason for the existence of  hyperbolic metamaterials is the support rapidly varying electromagnetic waves \cite{hyperlens1,hyperlens2} -- quadrupole transitions will be naturally excited. As a result, a  planar  hyperbolic metamaterial,  based on a quantum well superlattice,  operating at the  frequency of a forbidden transition, introduces the ideal platform to study such excitations.

\section{Electromagnetic response in the ``insurgent'' regime}

In this Letter, we will only consider the classical linear response regime. Without any assumptions on the coupling vs. isolation of the individual quantum wells in the hyperbolic metamaterial (the control of the well-to-well coupling is the key advantage of the digital alloys fabrication),  we can calculate the (single-electron) minibands
\begin{eqnarray}
\varepsilon_m\left({\bf k}\right),
\end{eqnarray}
and the corresponding electron single particle Bloch wavefunctions
\begin{eqnarray}
\psi_{m,{\bf k}}\left({\bf r}\right) & = &  \exp\left(i  {\bf k r}\right)\cdot  u_{m, k_z}\left(z\right), 
\end{eqnarray}
which offer the complete information for the calculation of the electromagnetic response of the metamaterial.

\subsection{Tangential permittivity}

The tangential-to-the-layers permittivity of the composite can be obtained using the standard Lindhard approach \cite{Lindhard} (see also \cite{E2N:PRA}), leading to
\begin{eqnarray}
\epsilon_\tau\left({\bf q}, \omega\right) &  = &  \epsilon_\infty + \frac{4 \pi e^2}{a q_\tau^2}
\sum_{{\bf k}_\tau; m} 
\nonumber \\
& \times &
\frac{  f_0\left(\varepsilon_{m,{\bf k} + {\bf q}_\tau/2} \right) - f_0\left(\varepsilon_{m,{\bf k}- {\bf q}_\tau/2} \right)  }{\varepsilon_m\left({\bf k}+ \frac{ {\bf q}_\tau}{2}\right) - \varepsilon_m\left({\bf k}-\frac{{\bf q}_\tau}{2}\right) - \hbar \omega - i \Gamma_{m}}, \ \ \ \ \ 
 \label{eq:eps_t}
\end{eqnarray}
where the subscript $\tau$ indicates the in-plane directions in the superlattice, so that
${\bf q}_\tau \equiv {\bf \hat{x}} q_x + \equiv {\bf \hat{y}} q_y $, and
\begin{eqnarray}
\sum_{{\bf k}_\tau; m}  & \equiv & \frac{1}{2 \pi^2} \sum_m \int d^2{\bf k}_\tau.
\end{eqnarray}

\subsection{Normal to the layers permittivity}

The electromagnetic response in the direction normal to the layers forming the metamaterial, requires a more careful consideration. 

To adequately represent the electric field
\begin{eqnarray}
{\bf E}\left({\bf r}, t\right) & = & {\bf \hat{z}}  \sum_{\bf q} {E}_{{\bf q}\omega} \exp\left(i {\bf q r} - i \omega t\right),
\end{eqnarray}
we need both the vector and the scalar potentials:
\begin{eqnarray}
\varphi\left({\bf r}, t\right) & = & \sum_{\bf q} {\varphi}_{{\bf q}\omega} \exp\left(i {\bf q r} - i \omega t\right), \\
{\bf A}\left({\bf r}, t\right) & = & \sum_{\bf q} {\bf A}_{{\bf q}\omega} \exp\left(i {\bf q r} - i \omega t\right), 
\end{eqnarray}
where
\begin{eqnarray}
{\varphi}_{{\bf q}\omega}  & = & i {E}_{{\bf q}\omega} / q_z, \label{eq:phi} \\
{\bf A}_{{\bf q}\omega}  & = & \frac{c}{\omega} 
\left(q_x {\bf \hat{x}} + q_y {\bf \hat{y}} \right) \ {\varphi}_{{\bf q}\omega}. \label{eq:A}
\end{eqnarray}

The effective single-particle Hamiltonian of the free carriers close to the band edge in the (coupled) quantum wells can be expressed as
\begin{eqnarray}
H_{\rm eff} & = & \frac{ \left({\bf \hat{p}} - \frac{e}{c} {\bf A}\left({\bf r},t\right) \right)^2 }{2 m_0}  + V\left(z\right)  + e \varphi\left({\bf r},t\right),
\end{eqnarray}
where $V(z)$ corresponds to the difference in the conduction band energies in different materials forming the electronic superlattice.

Using the standard electronic density matrix formalism, \cite{Boyd} for the charge density 
\begin{eqnarray}
\rho\left({\bf r}, t\right) & = & \sum_{\omega} \rho_\omega\left({\bf r}\right) \exp\left( - i \omega t\right)
\end{eqnarray}
in the linear response regime we obtain
\begin{eqnarray}
\rho_\omega\left({\bf r}\right) & = & e \sum_{m, n} 
\psi^*_{m,{\bf k}_m}\left({\bf r}\right) \  \psi_{n,{\bf k}_n}\left({\bf r}\right) \ \ \ \ \  \ \ \ \ \ 
 \nonumber \\
& \times & 
\frac{ f_0\left(\varepsilon_{n,{\bf k}_n} \right) - f_0\left(\varepsilon_{m,{\bf k}_m} \right)  }{\varepsilon_n\left({\bf k}_n\right) - \varepsilon_m\left({\bf k}_m\right) - \hbar \omega - i \Gamma_{mn}} \nonumber \\
& \times & 
\sum_{\bf q} \left[ \ 
{\varphi}_{{\bf q}\omega}  \ \langle n \left| e^{i {\bf q \cdot r}} \right| m \rangle
+ \frac{e {\bf A}_{{\bf q}\omega} }{2 m_0 c } \right. \nonumber \\
& \times & \left(  \langle n \left| \hat{\bf p} \ e^{i {\bf q \cdot r}} \right| m \rangle +  \langle n \left| e^{i {\bf q \cdot r}} \ \hat{\bf p}\right| m \rangle \right)
\bigg],
\end{eqnarray}
where $f_0$ is the Fermi-Dirac distribution function, and for any operator $\hat{A}$
\begin{eqnarray}
\langle n | \hat{A} | m \rangle \equiv \int d{\bf r} \ \psi^*_{n,{\bf k}_n}\left({\bf r}\right) \hat{A} \ \psi_{m,{\bf k}_m}\left({\bf r}\right),
\end{eqnarray}
leading to
\begin{eqnarray}
\rho\left({\bf r}, t\right) & = & \sum_{\omega} \rho_\omega\left({\bf r}\right) \exp\left( - i \omega t\right).
\end{eqnarray}
In the linear response regime we obtain
\begin{eqnarray}
\rho_\omega\left({\bf r}\right) & = & \sum_{{\bf q}; m, n} e^2 \left( {\varphi}_{{\bf q}\omega}  + 
\frac{\hbar {\bf A}_{{\bf q}\omega}  }{2 m_0 \omega }\cdot \left( {\bf k}_m + {\bf k}_n\right)  \right) \nonumber \\
& \times &
\psi^*_{m,{\bf k}_m}\left({\bf r}\right) \  \psi_{n,{\bf k}_n}\left({\bf r}\right) \  \langle n \left| e^{i {\bf q \cdot r}} \right| m \rangle
 \nonumber \\
& \times & 
\frac{ f_0\left(\varepsilon_{n,{\bf k}_n} \right) - f_0\left(\varepsilon_{m,{\bf k}_m} \right)  }{\varepsilon_n\left({\bf k}_n\right) - \varepsilon_m\left({\bf k}_m\right) - \hbar \omega - i \Gamma_{mn}}, 
\end{eqnarray}
where $f_0$ is the Fermi-Dirac distribution function, and for any operator $\hat{A}$
\begin{eqnarray}
\langle n | \hat{A} | m \rangle \equiv \int d{\bf r} \ \psi^*_{n,{\bf k}_n}\left({\bf r}\right) \hat{A} \ \psi_{m,{\bf k}_m}\left({\bf r}\right).
\end{eqnarray}

Neglecting the umklapp processes, \cite{Ziman} for the matrix elements we obtain
\begin{eqnarray}
\langle n \left| e^{i {\bf q \cdot r}} \right| m \rangle 
& = & 
\delta_{{\bf k}_n, {\bf k}_m + {\bf q}} \left(e^{i q_z z}\right)_{nm}, \\
\end{eqnarray}
where the integral in
\begin{eqnarray}
\left(e^{i q_z z}\right)_{nm} & \equiv & \int_{v_0} dz \ \psi^*_{n, {\bf k}_n}\left(z \right) \ e^{i q_z z} \ 
 \psi_{m, {\bf k}_m}\left(z \right) 
\end{eqnarray}
is taken over a single unit cell of the superlattice.

For the spatial Fourier component   of the unit cell - averaged charge density, 
\begin{eqnarray}
\rho_{{\bf q}\omega} & \equiv & \frac{1}{v_0} \int_{v_0} d{\bf r} \  \rho_\omega\left({\bf r}\right) \ 
\exp\left(- i{\bf q\cdot r}\right),
\end{eqnarray}
where $v_0$ is the metamaterial unit cell volume, we obtain
\begin{eqnarray}
\rho_{{\bf q}\omega} & = & e^2
\sum_{{\bf k}; m, n}  \left( {\varphi}_{{\bf q}\omega}  + 
\frac{\hbar   }{m_0 \omega }\ {\bf k}\cdot{\bf A}_{{\bf q}\omega}  \right)
\left| \left(e^{i q_z z}\right)_{mn}\right|^2
 \nonumber \\
& \times &
\frac{ f_0\left(\varepsilon_{n,{\bf k} + {\bf q}/2} \right) - f_0\left(\varepsilon_{m,{\bf k}- {\bf q}/2} \right)  }{\varepsilon_n\left({\bf k}+ \frac{\bf q}{2}\right) - \varepsilon_m\left({\bf k}-\frac{\bf q}{2}\right) - \hbar \omega - i \Gamma_{mn}}. 
\label{eq:rho_q}
\end{eqnarray}
Substituting (\ref{eq:A}) into (\ref{eq:rho_q}), for the $z$-component of the permittivity we obtain
\begin{eqnarray}
\epsilon_z\left({\bf q}, \omega\right) & \equiv & \epsilon_\infty - \frac{4 \pi}{q_z ^2} \frac{\rho_{{\bf q}\omega}}{\varphi_{{\bf q}\omega}} \ \ \ \ \ \ \ \ \nonumber \\
& = &  \epsilon_\infty + \frac{4 \pi e^2}{q_z^2}
\sum_{{\bf k}; m, n}  \left| \left(\exp\left({i q_z z}\right)\right)_{mn}\right|^2
\nonumber \\
& \times &
\frac{ \left( f_0\left(\varepsilon_{n,{\bf k} + {\bf q}/2} \right) - f_0\left(\varepsilon_{m,{\bf k}- {\bf q}/2} \right)  \right) }{\varepsilon_n\left({\bf k}+ \frac{\bf q}{2}\right) - \varepsilon_m\left({\bf k}-\frac{\bf q}{2}\right) - \hbar \omega - i \Gamma_{mn}} \nonumber \\
& \times & 
 \left(1 + \frac{\hbar \left( k_x q_x + k_y q_y \right)}{m_0 \omega}\right). 
 \label{eq:eps_z}
\end{eqnarray}
For high and/or thick barriers that suppress the inter-well tunneling, for isotropic free carrier dispersion in the bulk materials forming the superlattice, and in the metamaterial limit when in-plane electromagnetic wavenumber much smaller than in the inverse of the unit cell size,  $q_x,q_y \ll 1/a$,  the Eqn. (\ref{eq:eps_z}) reduces to
\begin{eqnarray}
\epsilon_z\left({\bf q}, \omega\right) & =  &
 \epsilon_\infty +  \frac{4 \pi e^2}{ q_z^2 a}
\sum_{{\bf k}_\tau; m, n}  \left| \left(\exp\left({i q_z z}\right)\right)_{mn}^{(0)}\right|^2
\nonumber \\
& \times &
\frac{ f_0\left(\varepsilon^{(0)}_{n} +\frac{\hbar^2 k_\tau^2}{2 m_0} \right) - f_0\left(\varepsilon^{(0)}_{m} +\frac{\hbar^2 k_\tau^2}{2 m_0} \right)  }{\varepsilon^{(0)}_n - \varepsilon^{(0)}_m - \hbar \omega - i \Gamma_{mn}}, 
\label{eq:e_zz}
\end{eqnarray}
where the subscript $\tau$ indicates the in-plane directions in the superlattice, and the matrix elements
\begin{eqnarray}
\left(e^{i q_z z}\right)^{(0)}_{nm} & \equiv & \int_{v_0} dz \ \phi^*_{n}\left(z \right) \ e^{i q_z z} \ 
 \phi_{m}\left(z \right) 
 \label{eq:ee_z}
\end{eqnarray}
are calculated using the  wavefunctions $\phi_m\left(z\right)$ of the single isolated quantum well forming the superlattice, with the corresponding energies  $\varepsilon_m^{(0)}$.

The description of the macroscopic electromagnetic response of a composite material in terms of the tensors of dielectric permittivity and magnetic permeability  relies on the relative small variation of the electromagnetic field on the scale of the metamaterial unit cell, \cite{BornWolf}
\begin{eqnarray}
\left| q_z \right| a < 1. \label{eq:qza}
\end{eqnarray}
 We can therefore expand the exponential $\exp\left(i q_z z\right)$ in the matrix elements of Eqn. (\ref{eq:ee_z}). In the leading order, 
\begin{eqnarray}
\left| \left(e^{i q_z z}\right)^{(0)}_{nm} \right|^2& = & q_z^2 \left| z_{mn} \right|^2 + { o}\left(q_z^2\right).
\label{eq:z}
\end{eqnarray}
However, in the case of a ``forbidden'' transition, due to the symmetry properties of the wavefunctions in an isolated 
quantum well, the matrix element $z_{mn}$ is exactly zero (e.g. in a symmetric quantum well, when both eigenstates involved in the transition, have the same parity). For this special ``forbidden'' case, in the leading order
\begin{eqnarray}
\left| \left(e^{i q_z z}\right)^{(0)}_{nm} \right|^2& = & \frac{q_z^4}{4} \left| \left(z^2\right)_{mn} \right|^2 + { o}\left(q_z^4\right).
\label{eq:z2}
\end{eqnarray}
We can combine (\ref{eq:z}) and (\ref{eq:z2}) into
\begin{eqnarray}
\left| \left(e^{i q_z z}\right)^{(0)}_{nm} \right|^2& = & q_z^2 \left[\left| z_{mn} \right|^2
+ \frac{q_z^2}{4} \left| \left(z^2\right)_{mn} \right|^2 \right].
\label{eq:z12}
\end{eqnarray}
Substituting (\ref{eq:z12}) into (\ref{eq:e_zz}), we obtain
\begin{eqnarray}
\epsilon_z\left({\bf q}, \omega\right) & =  & \epsilon_n\left( \omega\right) + \alpha\left(\omega\right) q_z^2,
\label{eq:eza}
\end{eqnarray}
where
\begin{eqnarray}
\epsilon_n\left( \omega\right)  & = &  \epsilon_\infty +  4 \pi e^2 n_0
\sum_{m, n}  \left| z_{mn} \right|^2 \nonumber \\
& \times &
\frac{ \hat{n}_{m}  - \hat{n}_{n}  }{\varepsilon^{(0)}_n - \varepsilon^{(0)}_m - \hbar \omega - i \Gamma_{mn}}, 
\label{eq:ezw}
\end{eqnarray}
and
\begin{eqnarray}
\alpha\left(\omega\right)  & = &
   \pi e^2 n_0
\sum_{m, n}  
\frac{\left| \left(z^2\right)_{mn} \right|^2 \left( \hat{n}_{m}  - \hat{n}_{n} \right)  }{\varepsilon^{(0)}_n - \varepsilon^{(0)}_m - \hbar \omega - i \Gamma_{mn}}. 
\end{eqnarray}
Here, the (total) electron density
\begin{eqnarray}
n_0 & \equiv & \frac{1}{a} \sum_{{\bf k}_\tau, m}  f_0\left(\varepsilon^{(0)}_{m} +\frac{\hbar^2 k_\tau^2}{2 m_0} \right)  \nonumber \\
& = & \frac{2}{a \left(2\pi\right)^2 } \int d{\bf k}_\tau  \sum_{ m}  f_0\left(\varepsilon^{(0)}_{m} +\frac{\hbar^2 k_\tau^2}{2 m_0} \right) \\
& = & \frac{m_0}{\pi a \hbar^2} \ \sum_{ m} \int_{\varepsilon^{(0)}_m}^\infty d\varepsilon  \  f_0\left(\varepsilon\right),
\end{eqnarray}
and the (dimensionless) factors $\hat{n}_m$ represent the relative fraction of the free carriers in the $m$-th subband:
\begin{eqnarray}
\hat{n}_m & = &
\frac{\int_{\varepsilon^{(0)}_m}^\infty d\varepsilon  \  f_0\left(\varepsilon\right)}{ \sum_{ \ell}  \ \int_{\varepsilon^{(0)}_\ell}^\infty d\varepsilon  \  f_0\left(\varepsilon\right)}.
\end{eqnarray}

Note that, due to the inequality (\ref{eq:qza}), the last term in Eqn. (\ref{eq:eza}) can only be comparable to the ``conventional'' contribution of $\epsilon_z\left(\omega\right)$ (see Eqn. (\ref{eq:ezw})) only close to the frequency of one of the forbidden transitions, $\overline{m} \leftrightarrow \overline{n}$,
\begin{eqnarray}
\omega \simeq \omega_0 \equiv \frac{\varepsilon^{(0)}_n - \varepsilon^{(0)}_m}{\hbar}.
\end{eqnarray}

In the regime, the contributions of all other (off-resonant) forbidden transitions, can be neglected, and
\begin{eqnarray}
\alpha\left(\omega\right) 
& \simeq & 
\frac{\Omega_0}{\omega_0 - \omega - i \gamma_0}, 
\label{eq:aw}
\end{eqnarray}
where $\gamma_0 \equiv  \Gamma_{\overline{m} \,  \overline{n}} / \hbar$, and
\begin{eqnarray}
\Omega_0 & \equiv & \frac{\pi e^2 n_0}{\hbar} \left| \left(z^2\right)_{mn} \right|^2 \left( \hat{n}_{m}  - \hat{n}_{n} \right). 
\end{eqnarray}

\section{Propagating waves in the ``insurgent" regime} 

With the dielectric permittivity of a hyperbolic  quantum well metamaterial at the frequency near a  classically forbidden 
transition,
\begin{eqnarray}
\epsilon\left({\bf q},\omega\right) & = &
\left(
\begin{array}{ccc}
 \epsilon_\tau\left(\omega\right) & 0 & 0 \\
 0 & \epsilon_\tau\left(\omega\right) & 0 \\
 0 & 0 & \epsilon_n\left(\omega\right) + \alpha\left(\omega\right) q_z^2 
 \end{array}
 \right),
\end{eqnarray}
for the wavevector of the propagating TM-polarized waves we obtain
\begin{eqnarray}
\frac{q_\tau^2}{\epsilon_n + \alpha q_z^2}
+
\frac{q_z^2}{\epsilon_\tau}
& = &
\frac{\omega^2}{c^2},
\end{eqnarray}
leading to 
\begin{eqnarray}
q_\tau & = & \pm \  \sqrt{\epsilon_n + \alpha q_z^2 } \ 
\sqrt{\frac{\omega^2}{c^2} - \frac{q_z^2}{\epsilon_\tau}}.
\end{eqnarray}

\begin{figure}[htbp] 
 \centering
   \includegraphics[width= 8.75 truecm ]{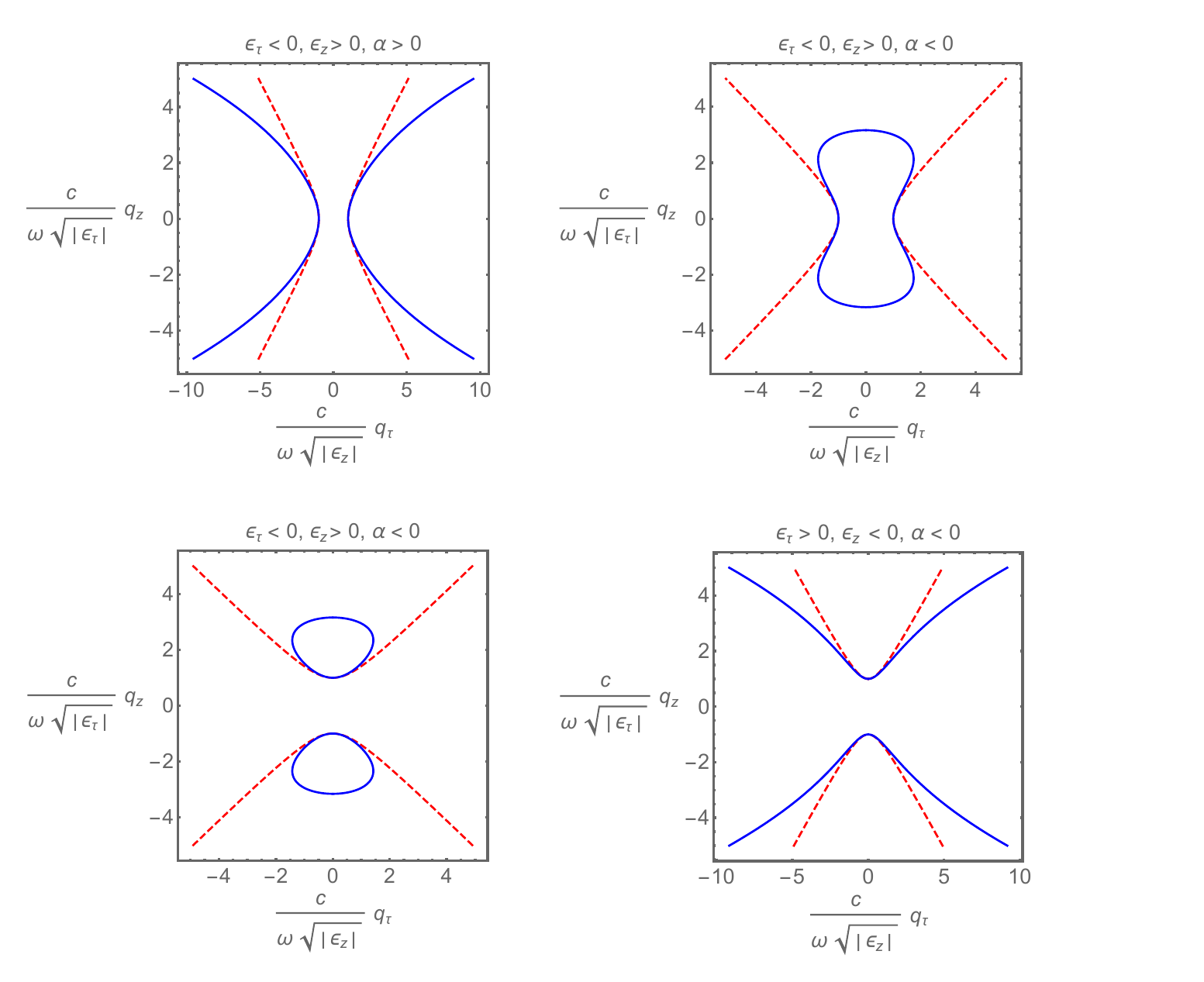} 
   \caption{ The iso-frequency curves for hyperbolic quantum well  metamaterial, with $\alpha \left| \epsilon_\tau / \epsilon_n \right| \left(\omega/c\right)^2 = \pm 0.1$  }
   \label{fig:iso_freq}
\end{figure}

The resulting iso-frequency curves in four possible  regimes corresponding to the different choices of the signs 
of $\epsilon_n$, $\epsilon_\tau$ and $\alpha$, are shown in Fig. \ref{fig:iso_freq}. There, a resonance in a forbidden transition always leads to a noticeable change in the momenta of the propagating waves, with the formation of additional waves when $\alpha < 0$.

In an experiment, the most striking effect of the resonant forbidden transition will be observed  in the regime 
${\rm Re}\left[ \epsilon_\tau \right] < 0$,  ${\rm Re}\left[ \epsilon_z \right] > 0$, $\alpha < 0$, when the ``metallic" reflectivity of a hyperbolic metamaterial is replaced by the effective ``dielectric'' behavior when the metamaterial 
now supports a propagating wave  -- even at the normal incidence. Direct measurements of the reflectivity and/or the transmission should clearly show a reduced reflection and enhanced transmission resulting from this ``insurgent'' dielectric response.

\section{Wave Equation in the ``insurgent" Regime, and the ABC problem} 

The presence of additional waves supported by an insurgent hyperbolic metamaterial, generally implies the need to the Additional Boundary Conditions -- which is generally referred to as the ABC problem \cite{ABC} that  received a considerable attention in late 1950s - early 1970s.\cite{ABC1,ABC2,ABC3} However, most of these (like e.g. imposing zero boundary condition for the normal-to-the-interface nonlocal component of the polarization \cite{ABC1}) are neither exact nor straightforward to implement in the conventional setting.

Instead, we will address this ABC-related problem here from the first principles. Assuming the coordinate-dependent tensors of the permittivity $\epsilon\left(z; \omega\right)$ and the spatial dispersion factor $\alpha\left(z; \omega\right)$, from the Maxwell's equations, for the TM-polarized field
\begin{eqnarray}
{\bf E}\left({\bf r}, t\right) 
& = &
\left(
{\bf \hat{x}}
E_x\left(z\right) + {\bf \hat{z}} E_z\left(z\right)\right) \  \exp\left(i q_x x - i \omega t\right), \\
{\bf B}\left({\bf r}, t\right) 
& = &
{\bf \hat{y}} B\left(z\right) \  \exp\left(i q_x x - i \omega t\right),
\end{eqnarray}
 we obtain
\begin{eqnarray}
& - & \alpha\left(z\right) \frac{\partial^3}{\partial z^3} \frac{1}{\epsilon_\tau\left(z\right)} \frac{\partial B}{\partial z}
+ \epsilon_z\left(z\right) \frac{\partial}{\partial z} \frac{1}{\epsilon_\tau\left(z\right)} \frac{\partial B}{\partial z}
\nonumber \\
& - &  \alpha\left( z\right) \left(\frac{\omega}{c}\right)^2 \frac{\partial^2 B}{\partial z^2}
 + \left(  \epsilon_z\left(z\right) \left(\frac{\omega}{c}\right)^2 - q_x^2 \right) B = 0.
 \ \ \ \ \label{eq:wave}
\end{eqnarray}
For a interface between different materials, corresponding to piecewise-constant functions  $\epsilon\left(z\right)$ and $\alpha\left(z\right)$, Eqn. (\ref{eq:wave}) implies the continuity of
\begin{eqnarray}
& {\rm (i)} & \ \ B\left(z\right), \label{eq:bc1} \\
& {\rm (ii)} & \ \frac{1}{\epsilon_\tau} \frac{\partial B}{\partial z}, \\
& {\rm (iii)} & \  \frac{\partial }{\partial z} \frac{1}{\epsilon_\tau} \frac{\partial B}{\partial z}, \label{eq:bc3} \\
& {\rm (iv)} &  \ \frac{\partial^2 }{\partial z^2} \frac{1}{\epsilon_\tau} \frac{\partial B}{\partial z} + \left(\frac{\omega}{c}\right)^2  \frac{\partial B}{\partial z}. \label{eq:bc4}
\end{eqnarray}

Note that, as $\alpha \to 0$, the wavenumbers of the new "additional waves'' when these are present, diverge, $\left|q\right| \to \infty$. In this limit, the boundary condition ({\ref{eq:bc4}) is enforced by the infinitesimal amplitude ($\sim \alpha^3$) of the additional wave with divergent wavenumber ($k \sim 1/\alpha$). As a result, in the special case of $\alpha = 0$ in one of the materials forming the interface, one only needs to consider the first three of the boundary conditions above, Eqns. (\ref{eq:bc1})-(\ref{eq:bc3}).

\section{Experimental Signatures of the Insurgent Response}

While the insurgent regime shows clear experimental signatures in all regimes, a particularly striking manifestation can
be observed for ${\rm Re}\left[ \epsilon_\tau\right] < 0$, ${\rm Re}\left[ \epsilon_n\right] > 0$, and ${\rm Re}\left[ \alpha\right] < 0$ (see Fig. \ref{fig:iso_freq}), corresponding to the formation of a new propagating mode in the ``metallic" regime of the ``bare'' hyperbolic metamaterial. In this case, direct coupling of the incident optical field to this mode leads to a dramatic reduction of the metamaterial reflectivity -- see Fig. \ref{fig:w}).

\begin{figure}[htbp] 
 \centering
   \includegraphics[width= 7.5 truecm ]{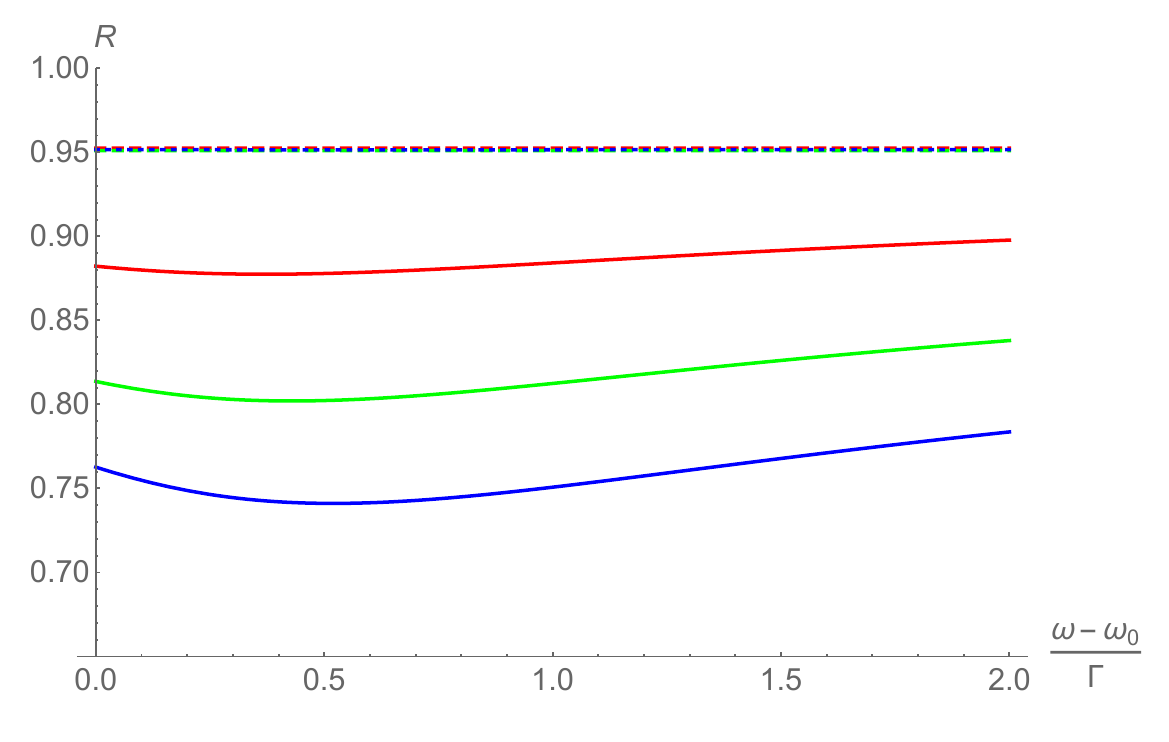} 
   \caption{ The reflectivity from the plane interface of the insurgent metamaterial with
   $\epsilon_\tau = -2$, $\epsilon_n = 3$ (solid lines) as a function of the light frequency $\omega$ (in units of the resonance width
   $\Gamma$), compared with the corresponding reflectivity from conventional hyperbolic metamaterial with the same permittivity tensor (dashed/dotted curves). Here, $\Omega_0 = 0.1 \Gamma$ (see Eqn. (\ref{eq:aw}))., and the angle of incidence $\theta = 30^\circ$ (red), $45^\circ$ (green) and $60^\circ$ (red).
   \label{fig:w}
  }
\end{figure}

This resonant reduction of the reflectivity also has a string signature in its angular dependence -- see Fig. \ref{fig:th}.

\begin{figure}[htbp] 
 \centering
   \includegraphics[width= 7.5 truecm ]{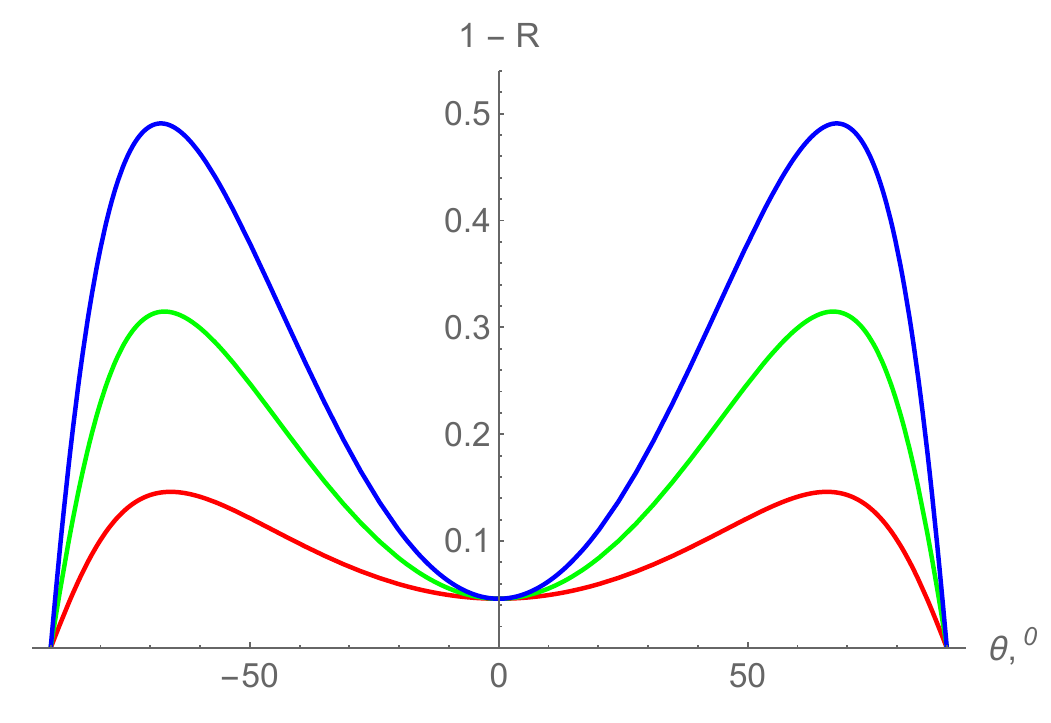} 
   \caption{ The difference of the insurgent metamaterial reflectance from unity, $1- R$, as a function of the angle of incidence $\theta$, for $\alpha = 0.01$ (red), $0.1$ (green) and $0.35$ (blue). The corresponding dielectric permittivity tensor components are $\epsilon_\tau = -2 + 0.1 i$ and $\epsilon_n = 3 + 0.1 i$.
 }
   \label{fig:th}
\end{figure}

\section{Acknowledgements}
The work was partially supported by the National Science Foundation (grant DMREF-2119157).
Author would like to thank Prof. Viktor Podolskiy for helpful discussions.

\end{document}